\documentclass[11pt, twoside]{article} 

\usepackage[round]{natbib}

\usepackage{url}

\usepackage{lineno}

\usepackage{graphicx}
\usepackage{amsmath}
\usepackage{url}

\usepackage{changepage}
\usepackage{booktabs}

\usepackage{multirow}
\usepackage{verbatim}
\usepackage{breqn}
\usepackage{enumitem}  

\usepackage{amsthm}
\usepackage{algorithm}
\usepackage{algorithmic}

\usepackage[text={14cm,20cm},centering]{geometry}

\title{Securing Tag-based recommender systems against profile injection attacks}
\title{Securing Tag-based recommender systems against profile injection attacks: A comparative study\footnote{This document is an extended pre-print of short paper published in \emph{RecSys 2018} by \cite{PitsilisRecSys18}. Please cite that version instead (see references).}\\ \Large{(Extended Report)}}

\author{Georgios K. Pitsilis, Heri Ramampiaro and Helge Langseth\\
\footnotesize Department of Computer Science\\
\footnotesize Norwegian University of Science and Technology\\
\footnotesize NO-7491 Trondheim, Norway\\
\footnotesize \texttt{georgios.pitsilis@gmail.com, heri@idi.ntnu.no, helge.langseth@ntnu.no} \\ }

\date{}

\begin{document}

\maketitle

\begin{abstract}
This work addresses the challenges related to attacks on collaborative tagging systems, which often comes in a form of 
malicious annotations or profile injection attacks. In particular, we study various countermeasures against two types of such attacks for social tagging systems, the Overload attack and the Piggyback attack. The countermeasure schemes studied here include baseline classifiers such as, Naive Bayes filter and Support Vector Machine, as well as a Deep Learning approach. Our evaluation performed over synthetic spam data generated from del.icio.us dataset, shows %
that in most cases, Deep Learning can outperform the classical solutions, providing high-level protection against threats.

\end{abstract}



\section{Introduction}
Recommender Systems (RS) are information filtering mechanisms, aiming at predicting the preference of users to particular resources.
\emph{Collaborative tagging}, a web-based service that is representative of the new Web 2.0 technology, allows users to store, share and annotate various kinds of web resources, such as news, blogs, and photos into social data repositories.
Also known as \emph{Social Tagging} systems, or \emph{Social Bookmarking} systems, such tools facilitate resource recommendations using the annotations of the users as input to various prediction algorithms.

With tags (a.k.a textual annotations) being a novel source of information, their concept has attracted the interest of scientists, with a literature rapidly expanding (\cite{Fernquist:2013,Heymann2008,Haustein2012}).
The open structure and the adaptive characteristics of RS have, however, made them vulnerable to various types of attacks.
In general, the aim of an attack to a RS is to promote or disapprove a product for the benefit of the attacker. 
A common way for an attacker to achieve this is to inject malicious profiles that correspond to fictitious identities into the system, in order to bias the system towards users being recommended a inferior product. 
Like other types of recommender systems, 
\emph{collaborative tagging} has attracted the interest of attackers, aiming to distort the system's behavior. Such an attack can be compared to attacking an ordinary rating-based Collaborative-filtering system in tag-based recommender systems.
The existence of these threats highlights the need for effective countermeasures for safeguarding such systems.

Our contribution in this paper is two-fold.
First, we provide a set of synthetic data, consisting of malicious annotations to be used for testing purposes. Second, we provide a comparative study of the effects of various attacks in a typical personalized tag-based RS. Along with this, we demonstrate the effectiveness of potential countermeasures against those attacks. Those include a \emph{deep learning} (DL) model adapted to the special requirements of the issue we come to address.

The rest of the paper is organized as follows.
In Section~\ref{preliminaries}, we describe in more detail the problem of attacks in tag-based recommender systems and describe the types of attack addressed in this work, in correspondence with two likely system scenarios. In Section~\ref{relWork}, we discuss the existing work in the field. 
In Section~\ref{Approach}, we present the classical and our proposed solution for tackling the issue; while in Section~\ref{Evaluation}, we describe the dataset that we used and the evaluation we performed, and discuss the results that we received. 
Finally, in Section~\ref{Conclusion}, we summarize our contributions and outline the future work.

\section{Preliminaries}
\label{preliminaries}
\subsection{Problem Statement}
\label{Statement}
In a social annotation system, the assignment of tags to resources by users is referred to as
\emph{folksonomies}, which are tuples of four components: $F=(U,T,R,A)$, where $U$,$T$ and $R$ are finite sets of users, tags and resources, respectively, and $A$ denotes a ternary relationship between them, e.g., $A\subseteq U\times T\times R$. A tag assignment $a=(u,t,r) \in A$ is an annotation of resource $r$ by user $u$ with tag $t$.

In general, an attack against a tagging system consists of one or more coordinated attack profiles, with each one being associated with a fictitious user identity.
The annotation history of such an identity is designed to bias the recommendation algorithm.
With this in mind, the research question that we address in this work is:

\begingroup
\addtolength\leftmargini{-0.2in}
\begin{quote}\it
How effective can be the various classifier schemes in the battle against the injection of bogus profiles, whose aim is to influence the personalized recommendations received by users in a tag-based recommender system?
\end{quote}
\endgroup

To answer this question, our main goals can be summarized as: 
\begin{itemize}
\item To study various classification schemes, including \emph{deep learning}, over known types of shilling attacks.
\item To evaluate the success of classification, as well as the actual effectiveness of filtering-out any bogus folksonomies into the recommendation process.
\end{itemize}

\subsection{Attacks Description}
\label{Attacks}
There are generally many commonalities between attacking a tag-based RS and the phenomenon of spamming. The main focus of this work is on special forms of attacks with interesting characteristics associated with a tag-based RS, as found in the literature~\citep{Ramezani2009}. To maintain consistency with the existing literature, the following terminology has been used:
\begin{itemize}
    \item \emph{Bogus} resource, is an item which an attacker wishes to promote, so that users' interest for this resource will increase.
    \item \emph{Popular} resource, is an item or a set of items with high visibility, which the attacker wishes to associate with the \emph{Bogus} resource.
    \item \emph{Popular} tag, is a tag of high visibility within the annotation system.
\end{itemize}

\subsubsection{Overload Attack}
As its name implies, the goal of this attack is to overload a tag context with a \emph{bogus} resource to achieve a correlation between the tag and that resource.
To accomplish this, an attacker would associate the \emph{bogus resource} with a number of \emph{popular} tags.
There are two variations of this attack, namely, \emph{Popular} and \emph{Focused}, with the latter targeting on a specific subset of users called \emph{likely buyers}, rather on all of them, as in the former. We restrict this work on the \emph{Popular} variation only.\\

\subsubsection{Piggyback Attack}
The objective of this attack is for the \emph{bogus resource} to ride the success of another highly reputable resource, which we call \emph{popular resource}. To accomplish this, an attacker associates the target resource with a highly reputable one, so that they would appear similar. 
A way to perform this attack is to apply a tag replication strategy of annotating the \emph{bogus resource}, using any \emph{popular tags} already associated with the \emph{popular resource}.

\subsubsection{System Operation Scenarios}
\label{SysOpScen}
To enhance the readers understanding about how the attacks could affect a tag-based recommender system, in this section we discuss two likely system-level scenarios for recommendation provision.
In the first one, we call \emph{tag-based resource recommendations}, the recommendations have no personalized form and are computed upon some input keyword provided by the user to a search query. 
In the second one, resource recommendations are personalized to users and are computed based on their profile similarity with the various resources. Also they can be provided in a proactive manner by the system.

\paragraph{Tag-based resource recommendations.}

In this scenario we assume that the resource recommendation algorithm receives a small set of tags as user's input query, and it recommends those resources which have the largest association with the tags in that query.

We present the following tagging scenarios, depicted in tables \ref{tab:overload_attack_example} and \ref{tab:piggyback_attack_example}.
For the shake of our example, there exist in total three resources in the RS (\emph{Resource \#1}, \emph{Resource \#2}, \emph{Resource \#3}), along with a set of tags chosen by various users for annotating each resource. The (\textbullet) mark denotes the choice of a user to annotate a particular resource with that tag.
Another two resources, marked as \emph{Bogus Resource \#1} and \emph{Bogus Resource \#2}, are used here as examples which the attacker wishes to promote, by applying the \emph{Overload} and \emph{Piggyback} attack respectively.
A number of bogus user profiles, are injected into the RS, marked in our example as, \emph{fake user A} and \emph{fake user B}, and used the most popular tags to annotate the \emph{Bogus} resource. 
For instance, the tag "\emph{car}" was included in the attack profiles as being one of the most frequently used in our example.

\begin{table*}[!htbp]
   \centering
   \caption{Overload attack example resource-tag matrix}
   \label{tab:overload_attack_example}
   \scalebox{0.9}{ 
   \begin{tabular}{c|c|c|c|c|c|c|c}
   \hline

   \multicolumn{2}{|c|} { \bfseries{\emph{Tag used}} $\rightarrow$ }   & dog    & food    & power   & car    &    web     &   \\
   \hline
   \hline
      \bfseries{Resource} $\downarrow$     & \bfseries{User} $\downarrow$   &          &          &         &          &              &   \bfseries{sum}     \\
   \hline
   \hline
   \multirow{4}{*}{\emph{Resource \#1}}     
   & \emph{Alice} &        &    \textbullet    &        &            &  &    \\
   \cline{2-8}  
   \multirow{5}{*}{}                  &  \emph{Bob} &          &   \textbullet      &   \textbullet     &        &        &        \\
   \cline{2-8}                        &  \emph{David} &    \textbullet     &          &         &   \textbullet    &   \textbullet    &        \\
   \cline{2-8}  
                                      &  \bfseries{sum}    &    1     &    2    &    1     &    1    &    1         &   6     \\
   \hline
   
   \multirow{3}{*}{\emph{Resource \#2}} &  \emph{Bob} &          &          &         &    \textbullet     &              &          \\
   \cline{2-8}  
   \multirow{5}{*}{}                  &  \emph{David} &    \textbullet     &    \textbullet     &         &          &              &        \\
   \cline{2-8}                        &  \bfseries{sum}    &    1     &    1     &    0    &    1     &       0       &   3     \\
   \hline   
             
   \multirow{3}{*}{\emph{Resource \#3}}  &  \emph{Alice} &          &          &   \textbullet     &         &      \textbullet       &          \\
   \cline{2-8}  
   \multirow{5}{*}{}                   &  \emph{Clark} &    \textbullet     &          &         &    \textbullet     &              &        \\
   \cline{2-8}                         &  \bfseries{sum}     &    1     &    0     &    1    &    1     &      1      &    4    \\
   \hline
   

   \multirow{3}{*}{\emph{Bogus Resource \#1}}  &  \emph{fake user A} &    \textbullet      &          &         &    \textbullet      &              &           \\
   \cline{2-8}  
   \multirow{5}{*}{}                   &  \emph{fake user B} &          &    \textbullet      &         &    \textbullet      &              &        \\
   \cline{2-8}                         & \bfseries{sum}     &    1      &     1     &    0     &     2     &      0        &        \\
   \hline
 
 
   
      \hline
   
   \end{tabular}}
\end{table*}

Before the attack, a prompt choice for any user who has queried the system using a popular tag, such as "\emph{car}", would be any of resources \emph{\#1}, \emph{\#2} or \emph{\#3}. That is because, this tag is the most frequently used for the above 3 resources.

For the case of \emph{Overload Attack}, an increase by the attacker to the frequency of an already \emph{Popular Tag}, has also the result of increasing the chances for the \emph{Bogus Resource \#1} to be presented as the most prompt choice in response to a user query.
In our example (see table \ref{tab:overload_attack_example}), suppose that after the attack, a user is querying the RS using the same tag (\emph{"car"}). This time he would be recommended by the system to choose \emph{Bogus Resource \#1} as the most prompt option.
The reason for that is simply because the tag "\emph{car}" after the attack has turned into the most frequently used annotation than before, thus becoming the most highly associated with the bogus resource.

\begin{table*}[!htbp]
   \centering
   \caption{Piggyback attack example resource-tag matrix}
   \label{tab:piggyback_attack_example}
   \scalebox{0.9}{ 
   \begin{tabular}{c|c|c|c|c|c|c|c}
   \hline

   \multicolumn{2}{|c|} { \bfseries{\emph{Tag used}} $\rightarrow$ }   & dog    & food    & power   & car    &    web     &   \\
   \hline
   \hline
      \bfseries{Resource} $\downarrow$     & \bfseries{User} $\downarrow$   &          &          &         &          &              &   \bfseries{sum}     \\
   \hline
   \hline
   \multirow{4}{*}{\emph{Resource \#1}}     
   & \emph{Alice} &        &    \textbullet    &        &            &  &    \\
   \cline{2-8}  
   \multirow{5}{*}{}                  &  \emph{Bob} &          &   \textbullet      &   \textbullet     &        &        &        \\
   \cline{2-8}                        &  \emph{David} &    \textbullet     &          &         &   \textbullet    &   \textbullet    &        \\
   \cline{2-8}  
                                      &  \bfseries{sum}    &    1     &    2    &    1     &    1    &    1         &   6     \\
   \hline
   
   \multirow{3}{*}{\emph{Resource \#2}} &  \emph{Bob} &          &          &         &    \textbullet     &              &          \\
   \cline{2-8}  
   \multirow{5}{*}{}                  &  \emph{David} &    \textbullet     &    \textbullet     &         &          &              &        \\
   \cline{2-8}                        &  \bfseries{sum}    &    1     &    1     &    0    &    1     &       0       &   3     \\
   \hline   
             
   \multirow{3}{*}{\emph{Resource \#3}}  &  \emph{Alice} &          &          &   \textbullet     &         &      \textbullet       &          \\
   \cline{2-8}  
   \multirow{5}{*}{}                   &  \emph{Clark} &    \textbullet     &          &         &    \textbullet     &              &        \\
   \cline{2-8}                         &  \bfseries{sum}     &    1     &    0     &    1    &    1     &      1      &    4    \\
   \hline
   
 
   
   \multirow{3}{*}{\emph{Bogus Resource \#2}}  &  \emph{fake user C} &    \textbullet      &    \textbullet      &         &    \textbullet      &              &           \\
   \cline{2-8}  
   \multirow{5}{*}{}                   &  \emph{fake user D} &          &    \textbullet      &         &    \textbullet      &              &        \\
   \cline{2-8}                         & \bfseries{sum}     &    1      &     2     &    0     &     2     &      0        &        \\
   \hline
 
   \hline   
   
      \hline
   
   \end{tabular}}
\end{table*}

For the \emph{Piggyback} attack, we assume that \emph{Resource \#1} is the most popular one, as soon as there is six tags in total (that is the largest number) that have been used for its annotation (indicated in \emph{sum} column in table \ref{tab:piggyback_attack_example}).
As per the attack's design, an intruder would apply a tag duplication strategy, aiming to associate his/her \emph{Bogus Resource \#2} with \emph{Resource \#1} (\emph{Target item}), in such a way that they would appear similarly annotated.
To achieve that, he would create a number of attack profiles (such as: fake users \emph{C} and \emph{D}), and annotate the \emph{Bogus Resource \#2}, using any popular tags already used for the annotation of the popular \emph{Resource \#1}, such as, \emph{"car"} or \emph{"food"}.
As a result of this, querying the recommender system with any popular tag associated with \emph{Resource \#1} (e.g "\emph{car}"), would result to \emph{Bogus Resource \#2} been recommended, rather \emph{Resource \#1}. 
That is because, a consequence of the attack is that "\emph{car}" becomes the most highly associated tag with the \emph{Bogus resource \#2} (annotated twice using that tag vs just once for \emph{Resource \#1}), and thus emerged into the prompt option.

\paragraph{Similarity-Based recommendations.}

In the second system-level scenario, the generated recommendations are entirely personalized, and are built upon the computed similarity between the users and the resources.

The two types of attacks  detailed in the previous paragraph can be adapted to the Similarity-based recommendation scheme as follows.
We take the previously mentioned scheme composed of five resources in total (non-fake and fake ones), namely \emph{Resource \#1, Resource \#2, Resource \#3} and \emph{Bogus Resource \#1, Bogus Resource \#2}, respectively, along with four legitimate users, \emph{Alice, Bob, Clark, David}, and two fakes ones, \emph{fake user C, fake user D}.
The objective of the attack in this scenario is the fake product to potentially become highly similar to normal users, thus emerging into a prompt option for them.

For computing such user-resource similarity, it is necessary that users' profiles and resources 
can be converted into vectors.
Various features, such as, the frequencies of tags derived from the annotation history of users, can serve in such case as the elementary information to compute the dot product of those vectors.
There is a plethora of algorithms suitable for expressing such vector similarity, such as, Eucledian~\citep{ONeill2006}, Jaccard~\citep{Hamers}, \emph{Vector Space}~\citep{Salton1975}, with the latter being the best suited to such requirement. 
Vector Space is a legacy recommendation algorithm for tag-based RS, which can be used for the provision of personalized resource recommendations to users~\citep{GEMMELL2012}.
This algorithm employs Cosine similarity which is a form of correlation that combines the tag frequencies. We refer to this model in more detail in section \ref{Evaluation} below.

To get some insight into the effects of the attacks in the above similarity-based scenario in accordance to the
example presented in tables \ref{tab:overload_attack_example} and \ref{tab:piggyback_attack_example}, we point out that while before the attack the prompt options for some legitimate users, such as, \emph{Bob} and \emph{Clark} would be very likely to choose \emph{Resource \#2} ($Sim[Bob,Res\#2]=0.668$ and $Sim[Clark,Res\#2]=0.820$) (as per eqn. \ref{eqn:cosine}),
for the same users, after the attack,  \emph{Bogus Resource \#1} becomes the top recommended option, due to its high similarity with them ($Sim[Bob , Bog\_Res\#1]=0.708$ and $Sim[Clark,Bog\_Res\#1]=0.868$ respectively).

Next, we refer to the best known solutions for tolerating attacks in RS, and then we present our approach along with a demonstration of the achieved performance in accordance to the Similarity-Based recommendation scenario.

\section{Related Work}
\label{relWork}

The issue of security in tag-based RS in not new, but it has so far been mainly approached by solutions associated with spam detection.
As per \cite{Heymann:2007}, anti-spam approaches in social networks are divided into 3 main categories: a) \emph{Prevention mechanisms}, such as CAPTCHAs \citep{VonAhn2003}, b) \emph{Rank-based} approaches which demote spam in search queries, and c) \emph{Identification-based} solutions which aim to detect and isolate any potentially threatening entities, such as a user or a resource. The focus of our work is on the latter type only.

To filter out non-legitimate users, various anti-spamming techniques have been developed by numerous researchers in the field, employing either a form of \emph{Bayesian} type filtering~\citep{Yazdani2012}, or other tag classification methods~\citep{Koutrika:2007}. Particular characteristics of the tags used in annotations are exhibited, such as the presence of suspicious ones into user profiles. In the above works the assumption made is that, any tags used by a legitimate user would coincide with those by other legitimate users.
Nevertheless, if attackers are aware of the correlations being used in an attack filtering policy, they might try to disguise their fake profiles by using non-suspicious tags~\citep{Koutrika:2007,Yazdani2012}. In such a case, any spam tag filtering is turned ineffective. 

\citet{Zhai2016} have developed a spam detection mechanism based on the user-behavior, that employs criteria such as the neighbors' honesty within the group, which the user is member of.
A similar filtering approach by \citet{Poorgholami} mixes features of tags and users together in the spam classification, that include the spamicity for tags from Baysian classification along with some user features derived from the social connectivity.

Neural network-based approaches, including deep learning (DL), are known to provide a good level of protection for the general task of spam detection in emails or social posts~\citep{gauri2017}.
%
However, despite the plethora of solutions existing today for spam classification, there is still a need for alternative and modern approaches, such as DL. 
%
%
%
%
%
To the best of our knowledge, DL has not been adequately investigated so far as a good solution to prevent attacks in tag-based RS, nor have quantifiable results been provided to demonstrate how the effectiveness of the countermeasures is reflected to the recommendation output.

\section{Securing the folksonomies}
\label{Approach}
In this section we briefly describe the three supervised learning classification algorithms that we included in our comparative study.

\subsection{Naive Bayes filtering}
Naive Bayes is known to be very effective for binary classification of emails in order to detect spam. 
To detect attacks, we applied Naive Bayes filtering for classifying folksonomies, based on the existence of tags in them. More specifically, we used the well-known formula derived from the Bayes theorem~\citep{Sahami98abayesian}.

For the case of detecting a fake folksonomy that was maliciously injected into the system, it works as follows.
For every folksonomy composed of $N$ tags we calculate the probability of being fake as:
$ p = \frac{1}{1+e^{n}}$, where 
$ n = {\sum_{i=1}^{N} [\ln(1-p_i)-\ln{p_i}]}$, in which: 
$p_i$ is the conditional probability that the folksonomy is fake, given that it contains tag $i$. 

\subsection{Support Vector Machine (SVM)}
Among the various classification schemes, SVM has gained increased popularity in the recent years, and its suitability for binary classification tasks \citep{Stanevski2005UsingSV} in NLP applications, is the main reason for the inclusion in this study.

We chose the \emph{linear kernel} function for SVM for separating the input folksonomies into two classes, \emph{legitimate} and \emph{malicious}.
The choice of that kernal function was driven by the fact that it gave the best results in the classification task, compared to other functions such as \emph{RBF}, \emph{Polynominal} and \emph{sigmoid}.
\emph{Squared Hinge} was the loss function used.
To apply this on our dataset, the input folksonomies were vectorized, before the classification. 
Vectorization is necessary for building a vocabulary of tags used in the folksonomies, so that each folksonomy is encoded as a fixed-length vector. 
Then, the tags in the vocabulary were transformed into TF-IDF values \citep{Salton:1988}.
TF-IDF expresses the importance of a tag based on how often it appears in that particular folksonomy, in relation to the whole dataset. 
In our case the values are computed by: $tfidf(t,d,D) = tf(td) \times idf(t,D)$, with $tf(t,d) = \frac{|d_t|}{|d|}$ and $idf(t,D) = ln\frac{D}{D_t}$, where $D$ denotes as the total number of folksonomies in the set, $D_t$ the number of folksonomies containing the tag $t$, $|d_t|$ the frequency of appearance of tag $t$  in folksonomy $d \in D$.
The above transformation was necessary in our case, to scale down the impact of the most frequent tags found in the dataset.

Speaking in terms of SVM classification, the challenge is to find a hyperplane of $(n-1)$ dimension for folksonomy data expressed in $n$ dimension feature space, so that to segregate the bogus from the legitimate class of folksonomies.
In our case, the number of features ($n$) is determined by the number of different tags exist in the corpus. 
The above challenge of finding a hyperplane is expressed by:
\begin{equation}
\sum_{i=1}^{n} w_i \cdot x_i + b_o = 0
\end{equation}
where:
$x_i$ in our case is a folksonomy expressed as element of n-dimentional feature space as: $\vec{x} = (x_1,...,x_n) \in {R}^n$,
$b_o \neq 0$, 
$w_i$ and $b_o$ are chosen based on the training data.
Based on that, elements $x_i \in R^{n}$ which reside above the hyperplane would satisfy: $\vec{b} \cdot \vec{x} + b_o > 0$, therefore they would be considered as members of the legitimate class, while those that satisfy $\vec{b} \cdot \vec{x} + b_o < 0$ would be the bogus ones.

\subsection{Deep Learning (DL)}
\begin{figure}
\centering
\hspace*{-0.1in}
\includegraphics[width=1.0\textwidth]{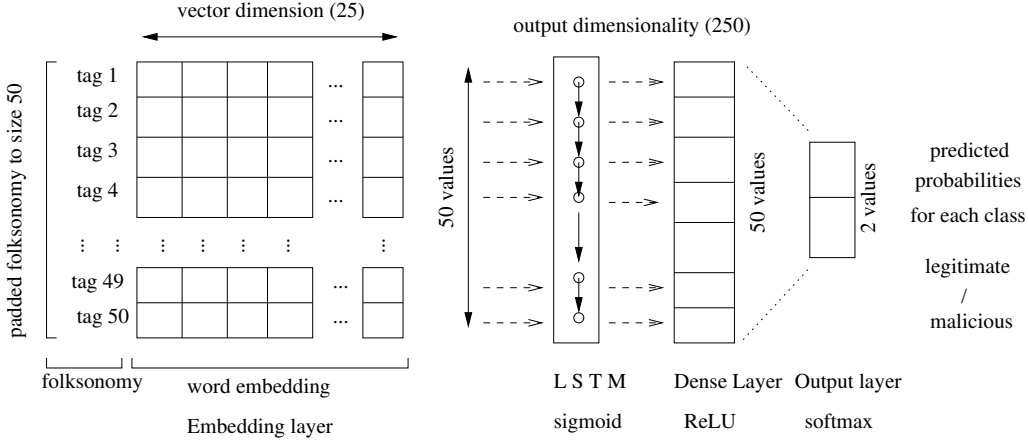}
\caption{The deep learning model architecture used}
\label{fig:model}
\end{figure}

We used a classification model, which employs a Long-Short-Term-Memory (LSTM)-based~\citep{Hochreiter:1997:LSTM}, Recurrent Neural Network (RNN) algorithm. This hybrid model has the strong benefit of working both with and without sequential data, and it is composed of four layers (see also Figure~\ref{fig:model}):

\textbf{\emph{i)} An \emph{input} layer.} 
 (a.k.a \emph{Embedding} layer), the size of which is associated with the size of the vocabulary of tags (corpus) used in the folksonomies, and it was set to 25. 
 
\textbf{\emph{ii)} A \emph{hidden} layer.},
 which is fully connected to the input and the subsequent layer. The dimensionality of the output space for this layer was set to 200 based on preliminary results. 
 
\textbf{\emph{iii)} A \emph{dense} layer}, used for improving the learning and stabilize the output, with size equal to that of the input folksonomy.
The ReLU activation function was chosen for this layer \citep{Hahnloser2000DigitalSA}, filtering the output as: $f(x) = x^+ = max(0,x)$.

\textbf{\emph{iv)} The \emph{output} layer}, which has 2 neurons to provide the classification output in the form of probabilities, for the Legitimate and Bogus class. For this layer, the \emph{softmax} activation function was chosen, with the second output value as such being the complement of the first one.

For embedding the input data into the model in a proper form, such as vectors, it requires applying the proper pre-processing onto the folksonomies, such as, word-based frequency vectorization.
That is, the tags in the dataset are indexed based on their frequency of appearance in the folksonomies.
In this way, a unique identifier is allocated to each tag and used as vector element to describe a folksonomy. 
For the case of \emph{del.icio.us} dataset we used in our experiment
we chose to apply a fixed vector size of 50 tags to represent folksonomies, which means, any smaller folksonomies had to be padded with zeroes to reach that size, while the longer ones where truncated to be no longer than 50 tags.

\section{Evaluation}
\label{Evaluation}

In order to demonstrate the effects of the attacks on the recommendation output, we
implemented the second system-level scenario we mention in section \ref{SysOpScen} for generating personalized, similarity-based recommendations. 
We focus on this scenario only, for two reasons. First, because it is more realistic, and second for its higher scientific importance.

For the evaluation we
employed the \emph{Vector Space} model for producing recommendations.
As the name implies, \emph{Vector Space} similarity is employed for computing the distance between a user's and an resource's vector, 
%
Each user is represented by a tag vector $u= [w_{t1}, w_{t2}, ..., w_{tn}]$ with $w_t$ denoting the weight of the particular tag $t$ on that user. Vector weights may be expressed through many ways, with the frequency of tags being the most common.
Likewise, each resource can be modeled as a vector $r= [u_{t1}, u_{t2}, ..., u_{tn}]$ over the same set of tags. 
Next, the user’s profiles and the resources can be matched over those tag expressions and compute the similarity value between them. 
\emph{Cosine Similarity} can be used to obtain these similarity scores between user profiles and rated resources. 
Then, by sorting the similarities in descending order we can eventually compile a \emph{top-k} list of personalized recommendations of resources for a specific user. 
The cosine of the two vectors in eqn.\ref{eqn:cosine} is derived from the Eucledian dot product formula, with $||\vec{t_n}||$ denote as the length of the vector $t_n$, and $\vec{t_u} \cdot \vec{t_r}$ is the inner product of a user vector $\vec{t_u}$ and a resource vector $\vec{t_r}$.

\begin{equation}
cos(\vec{t_1},\vec{t_2})=\frac{\vec{t_1} \cdot \vec{t_2}}{
||\vec{t_1}|| \ ||\vec{t_2}||}
\label{eqn:cosine}
\end{equation}

For our particular case, we further adapted the above model into the needs of folksonomy-based \emph{Social Tagging} systems.
First, we express every tag in the corpus in the form of a \emph{word2vec} vector~\citep{Mikolov2013}, extracted from a pre-trained set by \emph{Google}.
Next, for every folksonomy posted by a user we derive a single \emph{word2vec} representation by averaging the existing \emph{word2vec} values of all tags in that folksonomy.
Extending this to the user level, the \emph{word2vec} vectors of the folksonomies by that user can be averaged and finally receive a single vector representation of that user profile. The above formula can work similarly  for representing the resources as \emph{word2vec} vectors.
Finally, the similarity between a user and a resource can be worked out by computing the \emph{Cosine Similarity} value between their \emph{word2vec} vectors, that actually provides the distance that is necessary to know for compiling the personalized \emph{top-k} lists.
Computing the similarity on the \emph{word2vec} values was also found useful to avoiding issues related to cold-start users or resources. That is, any lacking of common tags in the annotation histories of a user and a resource would not make possible to compute the cosine similarity between them otherwise.  



\subsection{Dataset}
We chose to evaluate our hypothesis using annotation data from \emph{del.isio.us}\footnote{Del.isio.us dataset acquired from the work in \cite{basile_topical_2015}.}, a public dataset of folksonomies.
To reduce processing time, for every run
we selected randomly subsets of folksonomies corresponding to 3.000 users.
As opposed to other researchers who used pre-labeled datasets with spam data (e.g., \cite{Yazdani2012,Poorgholami,Zhai2016}), in our case, due to the absence of such bogus data we chose to evaluate over synthetic ones, we generated out of the original.
Moreover, this enabled us to study exclusive cases of attacks.

\begin{figure}[t]
   \centering
   \vspace{-3cm}
   \includegraphics[width=0.95\textwidth]{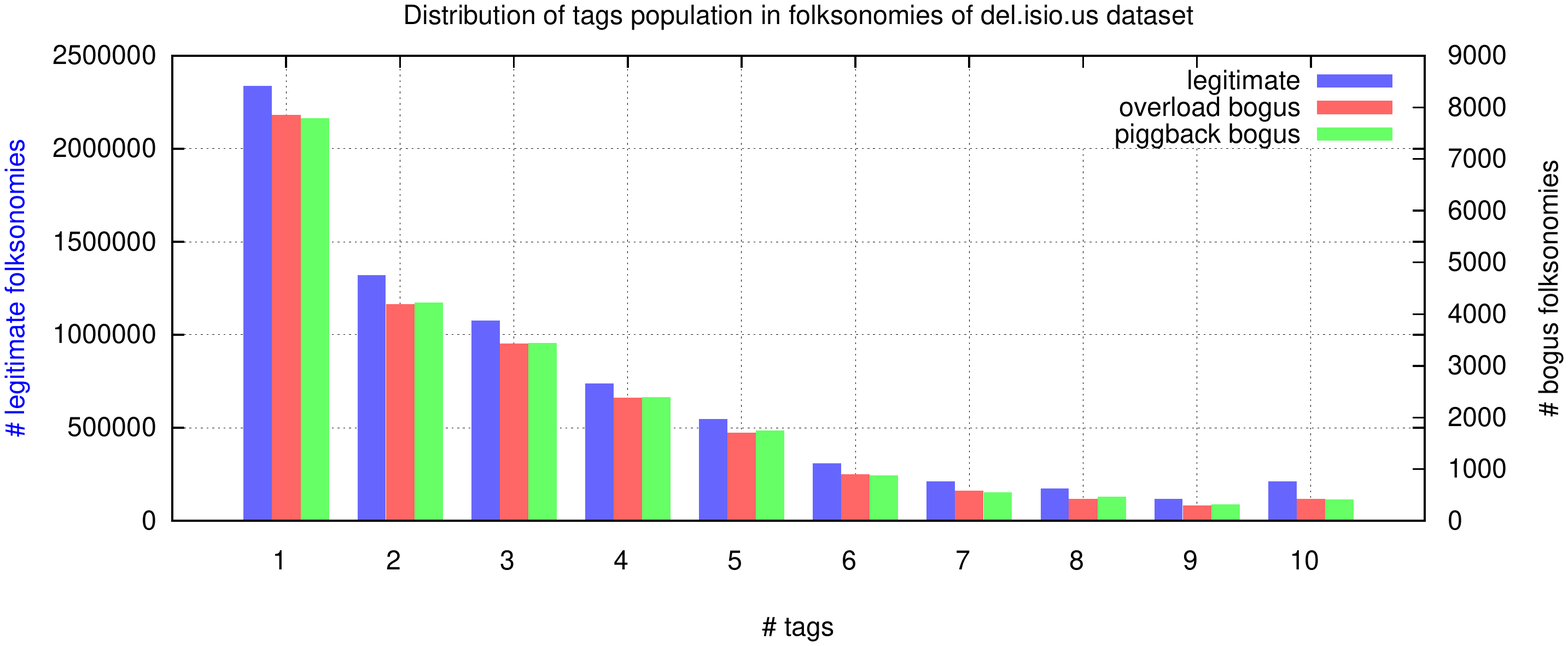}
   \vspace*{-0.3in}
   \caption{The distribution of tags in the original del.isio.us dataset along with the synthetic bogus data. $D_{KL}(P_{org}||Q_{ovl})=0.00481$ and $D_{KL}(P_{org}||Q_{pig})=0.00496$ where $P_{org},Q_{olv},Q_{pig}$ the legitimate, overload and piggyback distributions respectively.}


   \label{fig:distr}
\end{figure}

The size of the bogus folksonomies, as well as the tag selection, were determined according to the guidelines that we found in the literature~\citep{Ramezani2009} concerning the \emph{Overload} and \emph{Piggyback} attacks\footnote{The code as well as the dataset of bogus profiles will be available at the following address: https://github.com/gpitsilis/SecuringTagBasedrecSys}.
As such, to simulate the former, the tags of fake folksonomies were chosen out of the 75 most popular ones used in the legitimate folksonomies, while the max size of the fake folksonomies was limited to 50 tags.
The Popular tags were selected from those used for annotating the most popular resources.
To determine the actual number of tags to include in a fake folksonomy, we studied the distribution of the frequencies of tags in the legitimate folksonomies in the original dataset (\emph{del.icio.us}), and we build bogus ones with size that follows the same distribution.
In addition, for the \emph{Piggyback} attack, the tags we used for building the fake folksonomies were chosen out of those used in the 100 most highly annotated resources in the dataset. 
This above value was chosen as being suitable, given the sparsity of \emph{del.isio.us} dataset.
The distribution of tags in the original dataset, as well as in a randomly selected sample of synthetic fake folksonomies we used for the Overload and Piggyback attacks is shown in fig. \ref{fig:distr}. We computed the \emph{Kullback-Leibler} divergence value for discrete distributions of all classes, as per equation:
$D_{KL}(P||Q)=\sum\limits_{i} P(i) \log (\frac{Q(i)}{P(i)})$, by \cite{Kullback1987}. We used as input the normalized values of the frequencies of tags classes, and we noticed that there is indeed low divergence between the probability distributions of the original and the bogus folksonomies.

\begin{table}
   \centering
  \caption{Statistics of our \emph{del.isio.us} dataset.}
  \vspace{0.4cm}
   \label{tab:stat_dat}
   \scalebox{0.99}{ 
   \begin{tabular}{lc}
   \toprule
   \textbf{Statistics of dataset} & \textbf{\emph{del.isio.us}} \\
   \midrule
   \emph{\# folksonomies} & 73k \\
   \emph{\# users}        & 3k \\
   \emph{\# unique tags (corpus)}  & 42k \\
   \bottomrule
   \end{tabular}}
\end{table}

\subsection{Evaluation Metrics}
To demonstrate the impact of attacks, we chose to adopt a set of approved metrics from the literature \citep{Ramezani2009}. 

For the \emph{Overload} attack we used the following: 
\begin{itemize}
    \item \emph{F-Score} for the spam classification process. That is the harmonic mean of precision and recall, expressed as: $F = \frac{2 \cdot P \cdot R}{P + R}$. \emph{Precision} (P) is the ratio of the number of folksonomies correctly classified as belonging to the fake class, over the total number of folksonomies classified as fake, while \emph{Recall} (R), measures the ratio of folksonomies correctly classified as fake over the whole population of fake folksonomies.
    
    \item The \emph{Average rank} in which the \emph{bogus} resource appears in the users' \emph{top-k} lists of recommended resources.
    
    \item The \emph{population} affected by the attack, i.e., the number of users for whom the bogus resource has been recommended. That refers to whom the bogus resource has been included in their \emph{top-k} lists of personalized recommendations.
\end{itemize}

For the \emph{Piggyback} attack, the last metric refers to the population of users for whom the bogus resource has been ranked higher than the popular one in the \emph{top-k} recommendations they received.



\subsection{Experimental Setup}
For training the classifiers over the class of bogus data, a fix ratio of fake synthetic folksonomies was appended to the dataset, corresponding to  30\% of the total number of the legitimate ones. This value was chosen as being quite realistic for sufficiently training the classifiers and yet making possible to appropriately demonstrate the effects of the attacks.
The reason for choosing the same ratio of fake folksonomies allover the training process, was because we aimed to test the classification mechanism itself with regard to the size of bogus data been injected, rather than the effect of the bogus data on the training process.
As such, for testing each classification scheme, we chose to test over a variable size of attack.
This variable refers to the ratio of fake folksnomies injected into the system, over the total number of legitimate folksonomies, and it ranged from 0.1\% to 10\%.
Every experimental setup was run for 5 times, and the results were averaged.

To demonstrate the effectiveness of each classification algorithm, the task of producing \emph{top-k} recommendations was carried out both before and after applying a countermeasure in the classification process.
This means that, for each setup the synthetic fake folksonomies we generated, were mixed together with the legitimate ones from the original dataset, and then supplied into the \emph{Vector Space} model to compile the \emph{top-k} resource recommendations for each user.


The evaluation process can be explained as follows:
Let $S$ be the set of all folksonomies in the dataset, $L$ the originally legitimate, and $B$ the synthetically generated bogus ones, such that: $S = L \cup B$.
Let $f$ be the classifier to be evaluated receiving the above two sets $L,B$ as input. The classification output is composed of two sets $L_c,B_c = f(L,B)$, in which $L_c = L_L \cup B_L$ and $B_c$ = $L_B \cup B_B$. $L_L$,$B_L$ denote as the subsets of $L$ and $B$ respectively, which the classifier determined as "legitimate", and $L_B$ , $B_B$ denote as the subsets of $L$ and $B$ respectively, classified as "bogus".
Finally, the filtered output from the classifier $L_c$ is fed into the prediction algorithm to compute the accuracy of the classifier with respect to the contents of the \emph{top-k} lists determined by the \emph{Vector Space} algorithm.

For all three filtering algorithms we tested, we performed 10-fold cross validation over the dataset, in which the 9 folds were used for training and the remaining one for testing the algorithm on unseen data. 
The $k$ value that determines the size of the \emph{top-k} lists was set to 15.

We implemented the DL model in Keras toolkit \citep{chollet2015},
and used the {ADAM} optimization algorithm \citep{KingmaB14}. The \emph{categorical cross-entropy} was the learning objective chosen.
The size of the corpus, which is the number of different tags used in the sample folksonomies in our dataset, was set to 42k.

Furthermore, we chose to model the input folksonomy (see fig.~\ref{fig:model}), in the form of vectors using word-based frequency vectorization. In other words, the tags in the corpus were indexed based on their frequency of appearance in the corpus and the index value of each word was used as vector element to describe the folksonomy.
Finally, the DL model was allowed to run for a number of epochs until reaching an optimally trained state. That state is reached when the validation accuracy is maximized, while at the same time the error remains within $\pm 1\%$ of the lower ever figure within that fold throughout the cross validation.

\subsection{Results and Discussion}
We present the most interesting results of our experimentation.
As can be seen in Table~\ref{tab:Results_ov_del}, which shows the classifiers' performance expressed in \emph{F-score}, the DL approach outperforms the other approaches in classifying 
the bogus folksonomies for both attacks. The above metric refers exclusively to the ability of the classifier to detect the bogus profiles, and therefore it is independent of the size of attack.

\begin{table}[!htbp]
   \centering
   \caption{Classification accuracy for both attacks}
   \label{tab:Results_ov_del}
   \scalebox{0.9}{ 
   \begin{tabular}{c|c|c|c|c}
   \hline
   \textbf{Type of attack} &
   \multicolumn{1}{|c|} {\textbf{Classifier}  $\rightarrow$ }  & \textbf{SVM}   & \textbf{BAYES}  & \textbf{DL} \\
   \hline
   \hline
   \multirow{3}{*}{\emph{Overload}} & F-score (overall) &  0.9501 &	0.8339	& \bf{0.9570}   \\
   \cline{2-5}
   \multirow{3}{*}{}                & F-score (legit)  & 0.9665 & 0.9082 &	\bf{0.9709} \\
   \cline{2-5}                      & F-score (bogus)  & 0.8958 & 0.5863 & \bf{0.9104}  \\
   \hline
   \hline
   \multirow{3}{*}{\emph{Piggyback}} & F-scores (overall)       & 0.9680	 & 0.9009 &	\bf{0.9728}     \\
   \cline{2-5}
   \multirow{2}{*}{}                 & F-score (legit) & 0.97888 & 0.93878 & \bf{0.9818}  \\
   \cline{2-5}                       & F-score (bogus) & 0.9319	 & 0.7748 & \bf{0.9426}  \\
   \hline
   \end{tabular}}
\end{table}

%
%


\begin{figure}[!ht]
\centering
\includegraphics[width=0.95\textwidth]{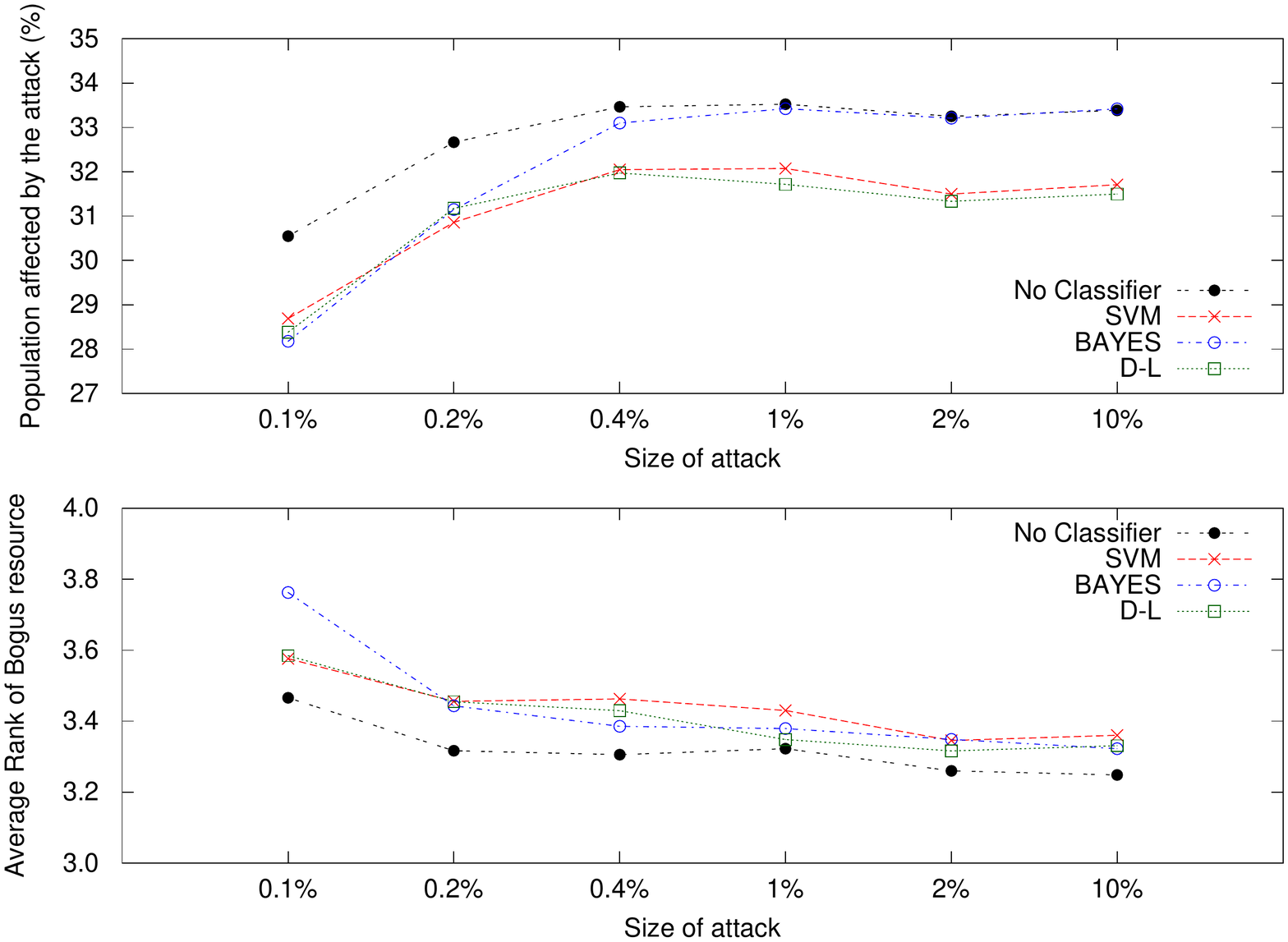}
\caption{Results for Overload attack showing the affected population (top - small values indicate strong resistance) and the rank of bogus resource (bottom - large values indicate strong resistance)}
\label{fig:graph_ol}
\end{figure}

\begin{figure}[!ht]
\centering
\includegraphics[width=0.95\textwidth]{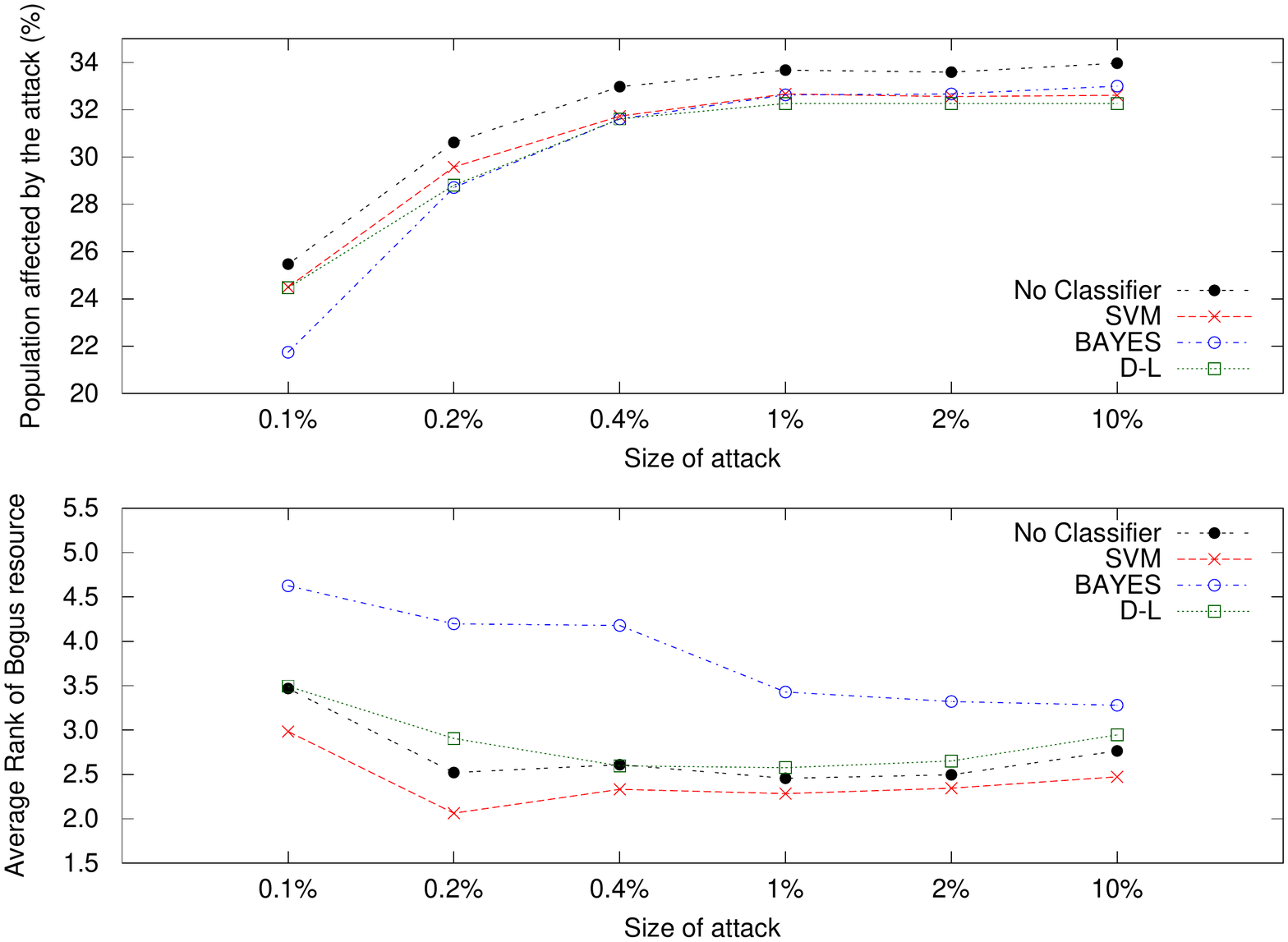}
\caption{Results for Piggyback attack showing the users population in which the bogus resource has ranked higher than the popular one (top - small values indicate strong resistance), and its actual rank (bottom - large values indicate strong resistance).}
\label{fig:graph_pg}
\end{figure}

\begin{figure}[!ht]
\centering
\hspace*{-0.42in}
\includegraphics[width=1.10\textwidth]{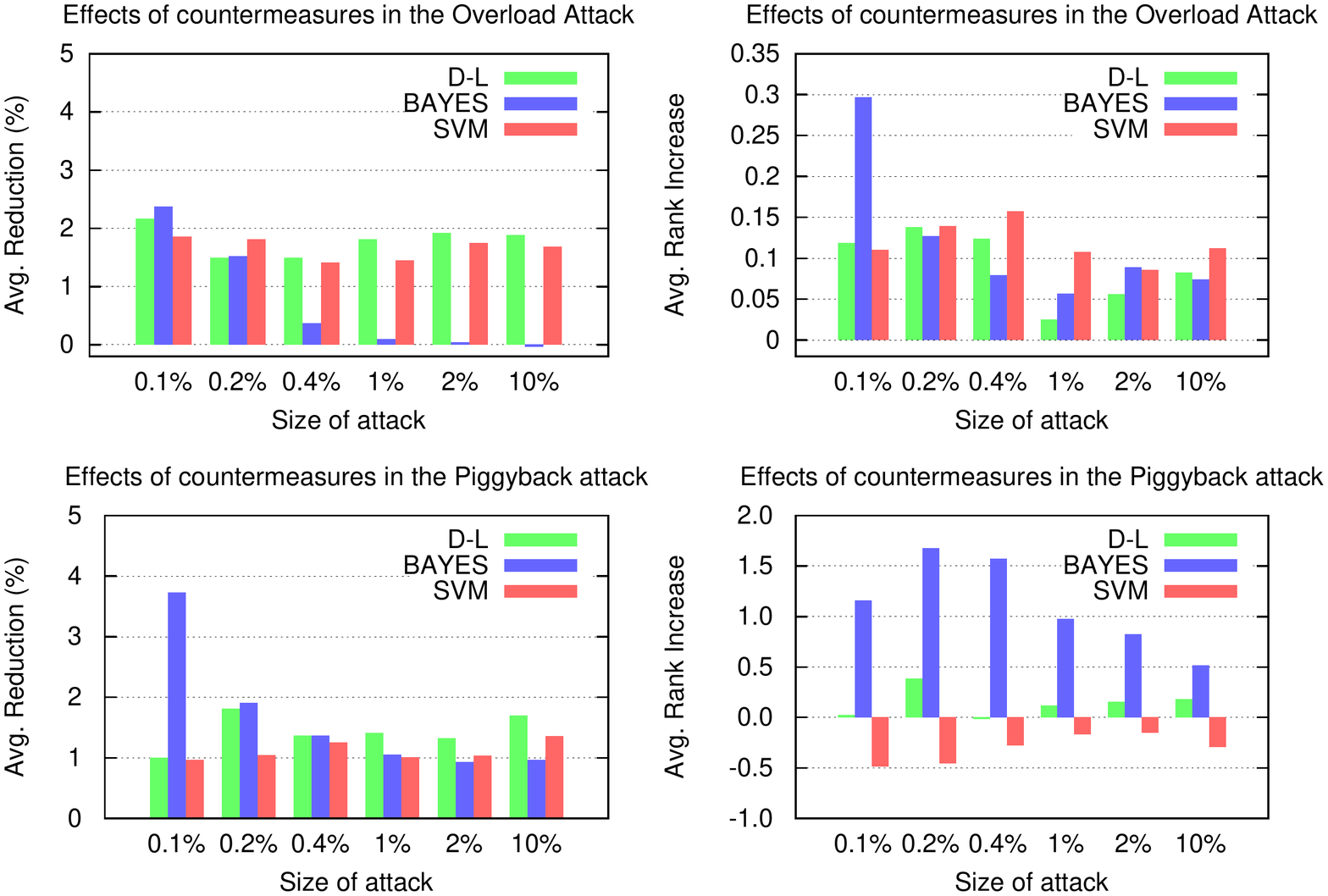}
\caption{Comparative results for both attacks, showing the gain of applying each countermeasure vs applying no classification policy, expressed in avg. reduction of affected population by the attack, and avg. rank increase of bogus item.}
\label{fig:graph_diff_ol}
\end{figure}

As far as the performance in the recommendation service itself, the population affected by the attack is shown in the top parts of Figures~\ref{fig:graph_ol} and \ref{fig:graph_pg}, while the rank of bogus resource, shown in the bottom parts of the same figures, refers to any users who have been affected by the attack.
As can be seen, very interestingly, even attacks of small scale are sufficient to render a significant population of users vulnerable.
In fact, the DL approach, in comparison to the other alternatives, provides in general, good resistance in preventing the intrusion of bogus resources into the users \emph{top-k} lists. 
That seems to be the case for both types of attacks.
Also interesting is the fact that in terms of the Avg. Rank of the Bogus resource, DL scales better for large sizes of attacks, in comparison to Bayes classifier, which seems to perform best for small attacks only.
(See fig.~\ref{fig:graph_ol}, bottom).
For the same metric in the piggyback attack, DL despite being the second best performing, the achieved performance improves with the increase of the attack size, as opposed to Bayes classifier, in which the performance is diminishing.

Comparing the effects of the Overload attack vs the Piggyback attack, we observe that in terms of the affected population, there is higher impact in the former attack, than in the second one. More specifically, as can be seen in the top sections of fig.\ref{fig:graph_ol} and fig.\ref{fig:graph_pg}, for \emph{No-classifier} results, the affected population ranges from 30.5 to 33.5 for the Overload attack, while for the Piggyback attack it ranges just from 25.9 to 33. This observation is in line with other researches in the field by \cite{Ramezani_2008}, even though expressed in different metrics.

To enhance understanding on the benefits received by the application of countermeasures, we present additional results in fig.~\ref{fig:graph_diff_ol}, showing the increase in the rank of the bogus item achieved by each classification scheme, as well as the average reduction in the affected population. The improvement by each classification method is expressed in relation to applying no countermeasures at all.

As opposed to our previous observation regarding the effects of the attacks, when it comes to the countermeasures, the Overload attack can be tolerated more easily than the Piggyback attack (see top parts of fig.~\ref{fig:graph_diff_ol}). The only exception is the Bayes classifier showing unstable behaviour.

In general, in terms of the increase of the rank of the bogus item in the \emph{top-k} lists as result to the application of countermeasures, we observe that in the Piggyback attack they are more effective, than they are on the Overload attack. The SVM classifier totally fails to provide any benefit over the application of no countermeasures for the piggyback attack as long as, on average, the bogus item appears at higher rank than the target item.

In conclusion, as we observed, the SVM and Bayes classifiers seem to behave best in one type of attack only, while DL being not the best in both, it yet provides a stable and fairly good resistance throughout any type and size of attack.

\section{Conclusions and Future Work}
\label{Conclusion}
In this paper we investigated the impact of spam filtering on resource recommendation in tag-based RS.
To simulate two known attacks, we generated synthetic fake folksonomies from original data, taken from \emph{del.isio.us} dataset.
%
Our experiments showed that the deep learning model outperformed all the legacy classifiers in terms of F-score and, 
in most cases, it can safeguard the user recommendations.
%
%

For future work, we plan to further improve the filtering performance by investigating other deep learning architectures, as well as experimenting with feature extraction from folksonomy data to feed the neural network. 
In addition, we intend to further generalize the results by exploring more resource annotation datasets, as well as performing extensive comparative studies.

\section{Acknowledgements}
This work has been financially supported by Telenor Research, Norway, through the collaboration project between NTNU and Telenor, and it has been carried out at the Norwegian Open AI-Lab. 

\bibliographystyle{ACM-Reference-Format}

\bibliography{tagging.bib}

\end{document}